\documentclass[%
 reprint,
superscriptaddress,
 amsmath,amssymb,
 aps,
]{revtex4-2} 

\usepackage{graphicx}% Include figure files
\usepackage{dcolumn}% Align table columns on decimal point
\usepackage{bm}% bold math
\usepackage{hyperref}% add hypertext capabilities
\usepackage[mathlines]{lineno}% Enable numbering of text and display math
\usepackage[subpreambles=true]{standalone}
\usepackage{import}

\begin{document}

\title[Fragile-to-strong transition in silica]{Understanding the fragile-to-strong transition in silica from microscopic dynamics}

\author{Zheng Yu}
\affiliation{Department of Materials Science and Engineering, University of Wisconsin-Madison, Madison, 53706, USA}
%  \altaffiliation[Also at ]{Physics Department, XYZ University.}%Lines break automatically or can be forced with \\
\author{Dane Morgan}
\affiliation{Department of Materials Science and Engineering, University of Wisconsin-Madison, Madison, 53706, USA}

\author{M. D. Ediger}
\affiliation{Department of Chemistry, University of Wisconsin-Madison, Madison, 53706, USA}

\author{Bu Wang}%
 \email{bu.wang@wisc.edu}
\affiliation{Department of Materials Science and Engineering, University of Wisconsin-Madison, Madison, 53706, USA}
\affiliation{%
 Department of Civil and Environmental Engineering, University of Wisconsin-Madison, Madison, 53706, USA
}%
% \author{\fnm{} \sur{}}

% \affil[1]{\orgdiv{Department of Materials Science and Engineering}, \orgname{University of Wisconsin Madison}, \orgaddress{\street{}, \city{Madison}, \postcode{53706}, \state{WI}, \country{USA}}}

% \affil[2]{\orgdiv{Department of Civil and Environmental Engineering}, \orgname{University of Wisconsin Madison}, \orgaddress{\street{}, \city{Madison}, \postcode{53706}, \state{WI}, \country{USA}}}

\begin{abstract}
In this work, we revisit the fragile-to-strong transition (FTS) in the simulated BKS silica from the perspective of microscopic dynamics in an effort to elucidate the dynamical behaviors of fragile and strong glass-forming liquids. Softness, which is a machine-learned feature from local atomic structures, is used to predict the microscopic activation energetics and long-term dynamics. The FTS is found to originate from a change in the temperature dependence of the microscopic activation energetics. Furthermore, results suggest there are two diffusion channels with different energy barriers in BKS silica. The fast dynamics at high temperatures is dominated by the channel with small energy barriers ($<\sim$1 eV), which is controlled by the short-range order. The rapid closing of this diffusion channel when lowering temperature leads to the fragile behavior. On the other hand, the slow dynamics at low temperatures is dominated by the channel with large energy barriers controlled by the medium-range order. This slow diffusion channel changes only subtly with temperature, leading to the strong behavior. The distributions of barriers in the two channels show different temperature dependences, causing a crossover at $\sim$3100 K. This transition temperature in microscopic dynamics is consistent with the inflection point in the configurational entropy, suggesting there is a fundamental correlation between microscopic dynamics and thermodynamics.
\end{abstract}

\keywords{fragile-to-strong, glass dynamics, softness, machine learning}



\maketitle

\section{Introduction}\label{sec1}

Glass-forming liquids based on the temperature dependence of dynamical slowing-down can be  classified into two groups.\cite{angellRelaxationLiquidsPolymers1991,angellFormationGlassesLiquids1995} If the viscosity (or other similar dynamic properties like diffusion coefficient and relaxation time) shows Arrhenius-like temperature dependence, e.g., in the case of silica, the liquid is referred to as ``strong". In other liquids including most organic and metallic glass formers, which are referred to as ``fragile", the slowing-down of the dynamics can be more drastic, with properties like viscosity showing super-Arrhenius temperature dependence. Fragility, which can be defined based on the degree of deviation from the Arrhenius behavior, is one of the most important concepts in glass physics, as it leads to intriguing questions on why dynamics slows down so quickly in some liquids and what controls the different behaviors.\cite{edigerSupercooledLiquidsGlasses1996} To answer these questions, the fragile-to-strong transition (FTS, or fragile-to-strong crossover) discovered in many glass-forming systems such as silica has attracted much attention.\cite{itoThermodynamicDeterminationFragility1999,saika-voivodFragiletostrongTransitionPolyamorphism2001,mallamaceTransportPropertiesGlassforming2010,zhangFragiletostrongTransitionMetallic2010} During FTS, a crossover in the liquid's dynamic behavior occurs without compositional changes or significant structural transformations. Understanding the cause of FTS could therefore provide unique insights into the origin of fragility.

The perspective on FTS so far is largely based on relating dynamics to thermodynamics, motivated by the Adam-Gibbs relation and further supported by the random first-order transition theory (RFOT).\cite{adamTemperatureDependenceCooperative1965,lubchenkoTheoryStructuralGlasses2007} The explanation focuses on the configurational entropy $S_c$ describing the number of inherent states, i.e., local minima, on potential energy landscape (PEL) explored by the system in equilibrium.\cite{goldsteinViscousLiquidsGlass1969,debenedettiSupercooledLiquidsGlass2001} Saika-Voivod et al. demonstrated based on molecular dynamics (MD) simulations that the Adam-Gibbs relation is obeyed in liquid silica, and that an inflection in $S_{c}$ may be responsible for the FTS.\cite{saika-voivodFragiletostrongTransitionPolyamorphism2001,saika-voivodFreeEnergyConfigurational2004} The work related this thermodynamic inflection to polyamorphism, a liquid-liquid phase transition, but this is not supported by experiments or simulations. Later, Saksaengwijit et al. explained the thermodynamic phenomenon by a depletion of inherent states below a cutoff on the silica PEL, inferred by MD simulations.\cite{saksaengwijitOriginFragiletoStrongCrossover2004} In this view, as the distribution of the sampled states touches the cutoff of the PEL, $S_{c}$ decreases more and more slowly with lowering temperature and eventually becomes constant. 

However, discrepancies still exist in the understandings of FTS. In our recent replica exchange molecular dynamics (REMD) simulations of silica,\cite{https://doi.org/10.18126/uffm-o07g} we did not observe a sudden depletion of states around the FTS temperature (i.e., from 3500 K down to 2000 K, as shown in Supplementary Information Fig. S2a). We attribute this discrepancy with the previous study to the ability in efficiently achieving equilibrium in REMD simulations, which is difficult in regular MD simulations of silica liquids below the FTS. Nonetheless, our simulation indeed confirmed the existence of an inflection point in $S_{c}$ at the FTS, as shown in SI Fig. S2b. The decrease of $S_{c}$ becomes slower as silica enters the strong region. However, because this can no longer be attributed to the sudden depletion of the states on the PEL, the source of this inflection point remains unclear.

In addition, the FTS may also be related to changes in the dynamics of local atomic rearrangements (hereinafter referred to as microscopic dynamics), which involves predominantly atomic hopping in the covalent network of silica. Saksaengwijit et al. observed the inherent structures above and below the FTS differ in the concentrations of short range defects.\cite{saksaengwijitOriginFragiletoStrongCrossover2004} This suggests that bond breaking, and therefore the microscopic dynamics, is more active above the transition. There have been debates on the contribution of microscopic dynamics to the dynamic slowdown in glass-forming liquids in general.\cite{wyartDoesGrowingStatic2017,berthierCanGlassTransition2019} Understanding the role of microscopic dynamics in FTS would provide direct insights to help resolve this issue.

Here, we investigate the features of the PEL related to microscopic dynamics, i.e., distribution of activation barriers, in BKS silica liquid across the FTS, by  means of machine learning (ML). Obtaining accurate statistics of microscopic activation barriers using traditional MD approaches would involve identifying large numbers of atom jumps (for both dynamically active and inactive atoms) and computing the associated barrier heights, which is very challenging computationally. Recently, Liu and colleagues developed a ML method that can successfully connect local atomic structure to microscopic dynamics.\cite{cubukIdentifyingStructuralFlow2015,schoenholzStructuralApproachRelaxation2016} The ML-generated quantity “softness”, which can be obtained solely from atomic structures, demonstrates clear correlations with dynamics in various systems.\cite{schoenholzRelationshipLocalStructure2017,sussmanDisconnectingStructureDynamics2017} Once the ML model is carefully trained and tested, microscopic activation barriers for individual atoms can be estimated from their local atomic environments based on their softnesses. By analyzing inherent structures collected from equilibrated MD simulations, we can examine how the barrier distribution changes with temperature in both fragile and strong silica liquids, thereby elucidating the role of microscopic dynamics in FTS.

\section{Results}\label{sec2}
\subsection{Predicting diffusion from local atomic structures}
The machine learning model is a key element of this study. Based on the methodology from Schoenholz et al.,\cite{schoenholzStructuralApproachRelaxation2016,cubukUnifyingFrameworkStrong2020} we implemented several modifications to improve the ML efficiency and accuracy for the silica system. Instead of using hundreds of symmetry functions to train the model, we only employ 10 structural features as inputs. The inputs include means and variances of distances and angles (listed in Sec. \ref{sec:methods_ML}) based on the tetrahedral orders of SiO\textsubscript{4} and SiSi\textsubscript{4} in silica. The output is whether an atom will rearrange in the next 1 ps. This is a smaller time window than previously used because we found it gives a better prediction accuracy. The ML training datasets are generated from inherent structures obtained from MD trajectories of equilibrated supercooled BKS silica liquid at 2600 K. After training, the accuracy of 86\% (and the recall, measuring the fraction of rearranging atoms that are correctly predicted, 79\%) was achieved, which is higher than previously obtained for the same system.\cite{cubukUnifyingFrameworkStrong2020} The ML quantity ``softness" is defined similarly to the previous studies, i.e., as the distance to the hyperplane for classification. Here we focus on Si atoms although we have obtained similar results were with O atoms as well. Details of the ML methods and the softness calculation can be found in Sec. \ref{sec:methods_ML} and SI.

\begin{figure}
	\centering
	\includegraphics[width=1.0\linewidth]{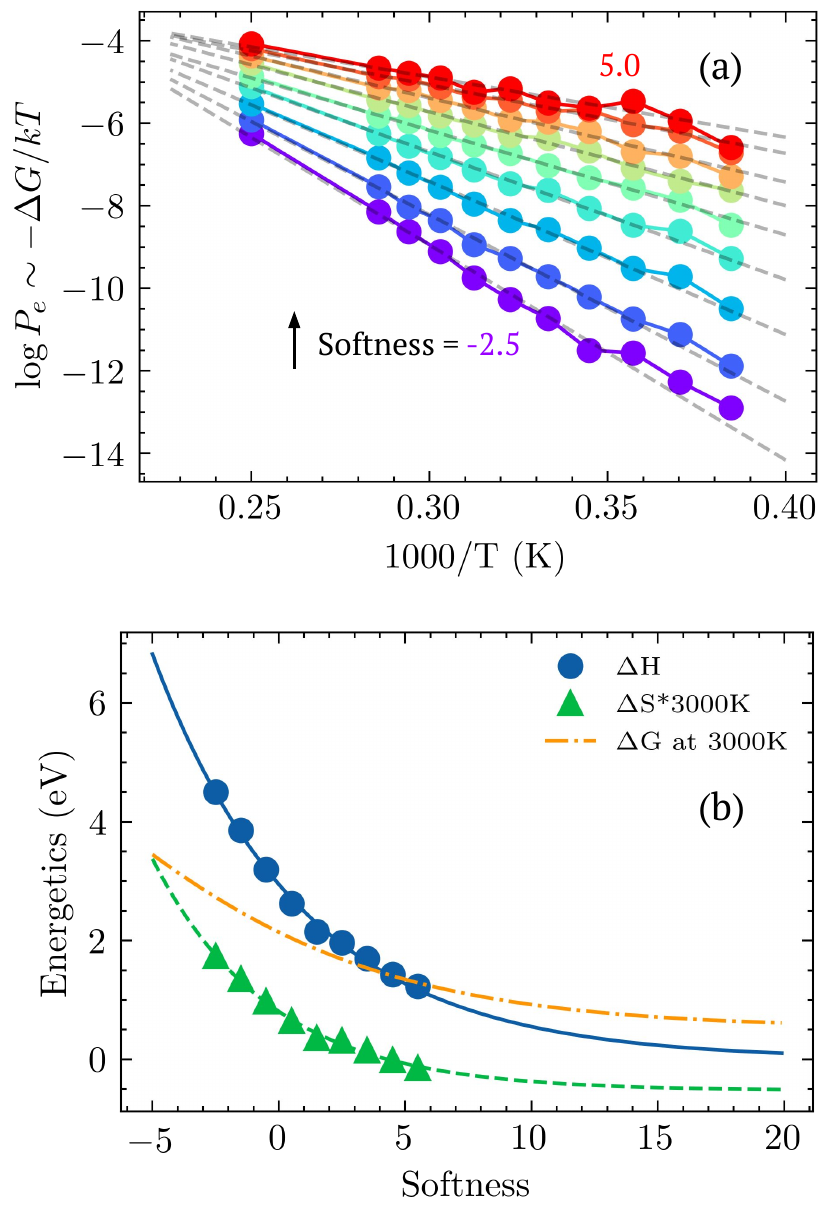}
	\caption{Application of the trained machine learning model to silica liquids at temperatures from 2600 to 4000 K. (a) Arrhenius plots of elementary hopping probability for atoms with different softnesses. The arrow points to increasing softness in different lines from -2.5 to 5.0. (b) Enthalpy, entropy, and free energy of activation as functions of softness. The lines are exponential fittings detailed in the main text.
	}
	\label{fig:softness_dynamics}
\end{figure}

After training, the ML model is utilized to study the FTS in two steps. In the first step, we use softness to estimate microscopic dynamics and the associated energy barrier statistics. The relationship between softness and activation energy of microscopic dynamics is established by applying the ML model to various temperatures from 2600 to 4000 K and statistically computing the rearrangement probability as a function of softness (denoted as $s$). For the selected time window of 1 ps, the total rearrangement probability, $P_R$($t$=1 ps) can be expressed as a function of an elementary rearrangement probability $P_e$ (or atomic hopping probability) by
\begin{equation}
	P_R(t<<\tau_{\alpha}) = 1-(1-P_e)^n
\end{equation}
where $n$ is the number of hopping attempts within $t$. $n$ within the temperature range we investigate is assumed as 30 for 1 ps, based on the vibrational density of states of amorphous silica.\cite{benoitVibrationalDynamicsVitreous2002,bhattaraiVibrationsAmorphousSilica2016} The previous study used $n=1$ for this value.\cite{cubukUnifyingFrameworkStrong2020} Since the hopping probability is small, choosing different $n$ values mainly affects the absolute value of $P_e$ we obtain rather than its trend with temperature. The elementary rearrangement probability $P_e$ can be described based on the transition state theory as
\begin{equation}\label{eq:eq2}
	P_e(T,s) = \exp(-\frac{\Delta G(T,s)}{k_B T})=\exp(\frac{\Delta S(s)}{k_B})\exp(-\frac{\Delta H(s)}{k_B T})
\end{equation}
where $\Delta G$, $\Delta S$, and $\Delta H$ are free energy, entropy, and enthalpy of activation, respectively. Note that the small variation in the hopping attempt frequency over the temperature range is accounted for through the prefactor term of $\Delta S$. As shown in Fig. \ref{fig:softness_dynamics}a, the rearrangement probability for atoms with a given softness shows an Arrhenius behavior with temperature. The slope of the Arrhenius function gives $\Delta H$ and the intercept $\Delta S$. Fig. \ref{fig:softness_dynamics}b summarizes the calculated $\Delta H$ and $\Delta S$  as functions of softness $s$. The enthalpy and entropy of activation both decrease with increasing softness. This relationship can be fitted with exponential function  $\Delta H = A\exp(-B\cdot s)$, where $A$ and $B$ are fitting parameters. Because a minimum value of zero is expected for $\Delta H$ (i.e., zero barrier height) but not for $\Delta S$ (which is simply the entropy of the transition state with respect to the ground state), a constant is included to the exponential fitting for $\Delta S$, $\Delta S = A\exp(-B\cdot s)+\Delta S_0$. Nevertheless, the activation free energy can now be established as a function of softness. Combined with softness distributions obtained by applying the ML model to inherent structures generated from MD simulations, we are able to predict microscopic dynamics and associated activation energy distribution for silica liquids at various temperatures. 

\begin{figure}
	\centering
	\includegraphics[width=1.0\linewidth]{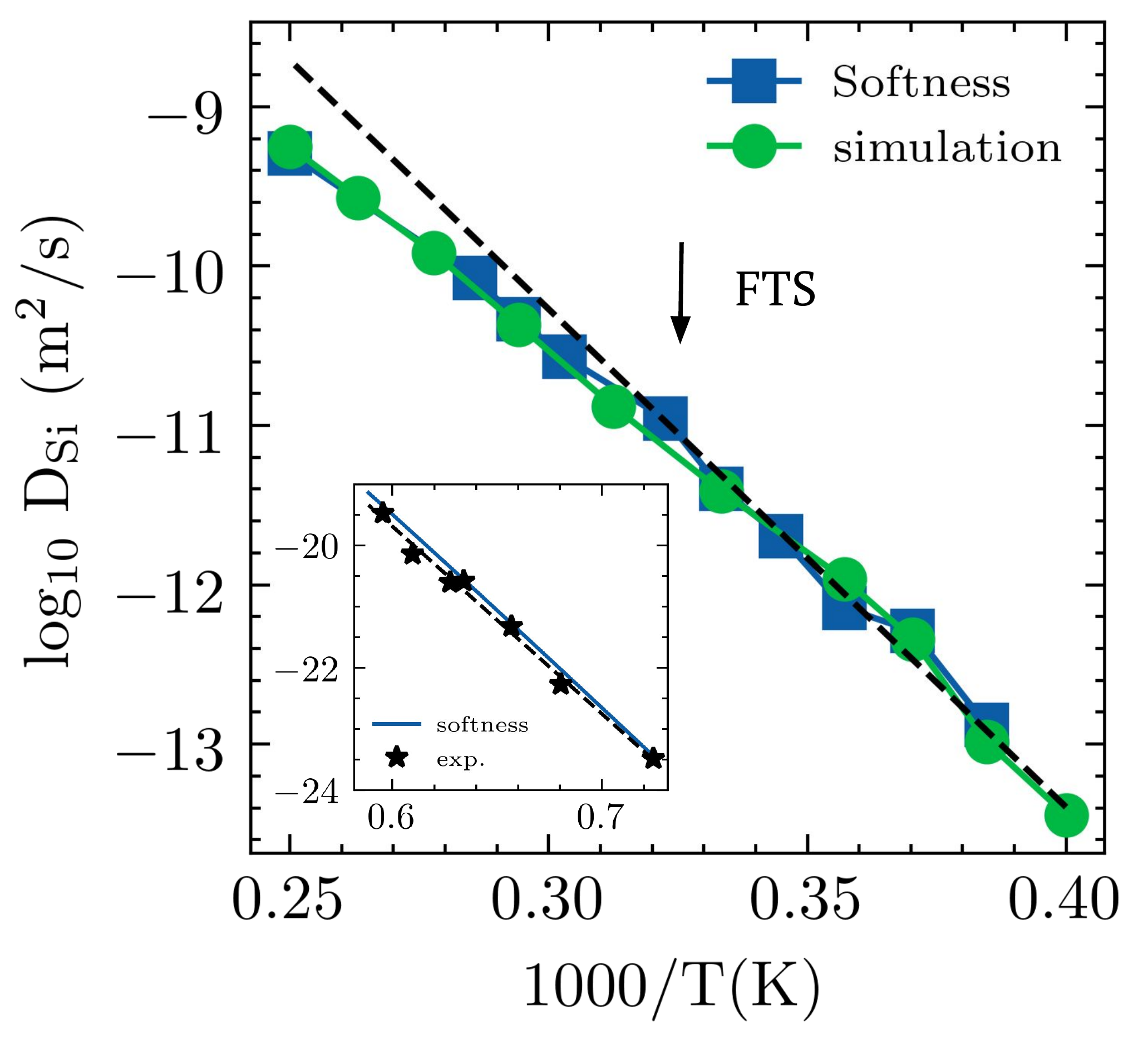}
	\caption{Diffusion coefficients of Si atoms predicted by softness compared to those calculated directly from MD simulations. The diffusion coefficient is found approximately proportional to the square of the elementary hopping probability. Details of this relationship are discussed in SI. The dashed line is an Arrhenius fitting of the low-temperature data. An FTS is clearly shown in the predicted diffusion coefficients around 3100 K, as noted by the arrow. The inset shows the diffusion coefficients around the experimental $T_g$ predicted by extrapolating softness distribution to lower temperatures (detailed in SI Sec. S4) are in good agreement with experimental data.\cite{brebecDiffusionSiliciumDans1980a}}
	\label{fig:D}
\end{figure}

In the second step, we investigate FTS using softness based on the correlation between short- and long- term dynamics in silica liquids. The  elementary hopping probability predicted so far is linearly related to short-term dynamics, represented by the mean squared displacements (MSD) over a short time relative to the relaxation time, e.g., 1 ps, assuming a constant hopping distance $R$:
\begin{equation}
	\mathrm{MSD}_{\mathrm{IS}}(t<<\tau_{\alpha}) \propto \langle R^2 \exp(-\frac{\Delta G}{k_B T}) \rangle_{eq} \propto R^2 \langle \exp (-\frac{\Delta G}{k_B T}) \rangle_{eq} 
\end{equation}
Note that the short-term dynamics here is based on inherent structures and therefore the thermal vibrations have been excluded. As shown in Fig. S4a, MSD\textsubscript{IS}(1ps) predicted by softness agree well with those calculated directly from MD trajectories. However, the FTS in silica liquid manifests in long-term dynamics, e.g., in bulk diffusion coefficients. To bridge the dynamics at different time frames, we found that the bulk diffusivity in silica liquid is connected to the elementary hopping probability (or short-term/microscopic dynamics) through a quadratic relationship:
\begin{equation}
	D \propto \langle \exp(-\frac{\Delta G}{k_B T}) \rangle_{eq}^\alpha 
\end{equation}
where $\alpha$ is found to be around 2 based on  MD simulation results. This relationship is phenomenological and based on the evolution of the time dependence of MSD from short- to long- term. The details of the calculations are reported in SI. Extrapolating short-term dynamics to long term using this relationship, we can now estimate bulk diffusivity using softness. As shown in Fig. \ref{fig:D}, diffusion coefficients of Si atoms predicted by softness are in good agreements with those calculated directly from 300 ns MD trajectories. Similar to the diffusion behavior directly observed in MD simulations, a FTS is clearly present in the softness-predicted diffusion coefficient around 3100 K.  By demonstrating the FTS can be predicted from local structures that control atomic hopping, we show that the transition may be rooted in microscopic dynamics that is directly resolved by softness.

\subsection{Activation energetics of microscopic dynamics}
\begin{figure*}
	\centering
	\includegraphics[width=1.0\linewidth]{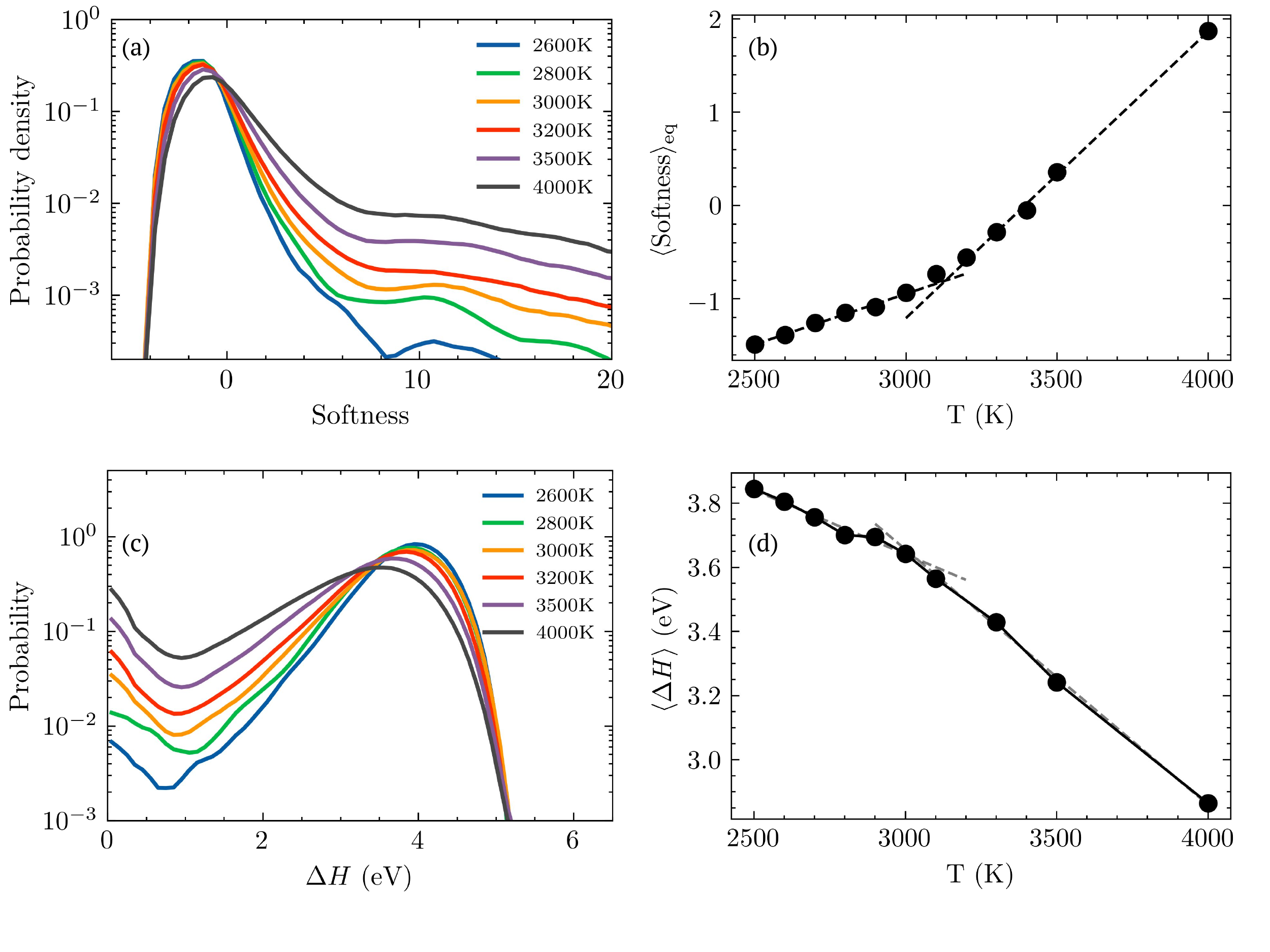}
	\caption{(a) Softness distributions at different temperatures. (b) Average softness as a function of temperature. The subscript `eq' denotes that the structures used for softness statistical analysis are all in equilibrium at the corresponding temperature. (c) Bimodal distributions of activation enthalpy of microscopic dynamics at different temperatures. (d) Average activation enthalpy of microscopic dynamics as a function of temperature, showing a kink at $\sim$3100 K corresponding to the FTS. }
	\label{fig:distributions}
\end{figure*}

We now utilize the ML model to further investigate the origin of the FTS. Taking atomic snapshots of the silica liquid, atoms in different local environments have different softnesses, representing their propensity for different microscopic dynamics. The softness distributions in liquid silica at different temperatures are shown in Fig. \ref{fig:distributions}a. At the highest temperature we investigated, 4000 K, the softness distribution shows a peak around -1 with a long tail of large softness extending to $s>20$. At this temperature, atoms with large softnesses, i.e., that can easily hop or rearrange, pervade in the liquid. The tail of large softnesses shrinks in proportion to the main peak as the temperature decreases. At 2600 K, most atoms show low propensity for local rearrangement, i.e., covered under the main peak. The distinct peak around -1 present at all temperatures suggests there is a main type of atomic environments in silica liquids associated with slow microscopic dynamics. Our ML results indicate this corresponds to the ideal tetrahedral order in the short range. Atoms without such ordered atomic environments, captured by the long tail of large softnesses, contribute to fast microscopic dynamics. The relative proportions of the two groups change with temperature. As such, when the averaged softness in the liquid is plotted versus temperature in Fig. \ref{fig:distributions}b, two linear portions can be observed with a break of slope at $\sim$3100 K, coinciding with the FTS.

We further predict the distributions of microscopic activation enthalpies $\Delta H$, i.e., the energy barriers for the elementary atomic hops, based on the relationship in Fig. \ref{fig:softness_dynamics}b. Distributions of the energy barriers at all temperatures are clearly bimodal, as shown in Fig. \ref{fig:distributions}c, with one main peak at $\sim$4 eV and a smaller peak around zero. Again, the main peak corresponds to atoms in well-coordinated tetrahedral environments. Hopping of these atoms is associated with substantial energy barriers. As the temperature decreases, the proportion of these atoms and associated energy barrier height both increase slowly, as suggested by the main peak increasing in intensity and shifting right. The changes in this peak become even more subtle below the FTS, suggesting the coordination environment becomes rather stable for these atoms at low temperatures. The secondary peak around $\Delta H\sim0$ represents a microscopic dynamics channel associated with very small energy barriers. With lowering temperature, the height of this peak decreases quadratically in the case of liquid silica (please refer to SI for more detailed analysis).\cite{yuStructuralSignaturesThermodynamic2021} As a result of different temperature dependencies of the two microscopic dynamics channels, the averaged activation enthalpy shows a transition around 3100 K, as shown in Fig. \ref{fig:distributions}d. Note that, although the overall activation energy associated with the bulk diffusivity is not equal to the averaged microscopic activation enthalpy, they are linearly related, as shown in Fig. S6a. The activation energy in the low temperature region changes slowly with temperature, leading to a stronger behavior than the high temperature region.

\begin{figure*}
	\centering
	\includegraphics[width=1.0\linewidth]{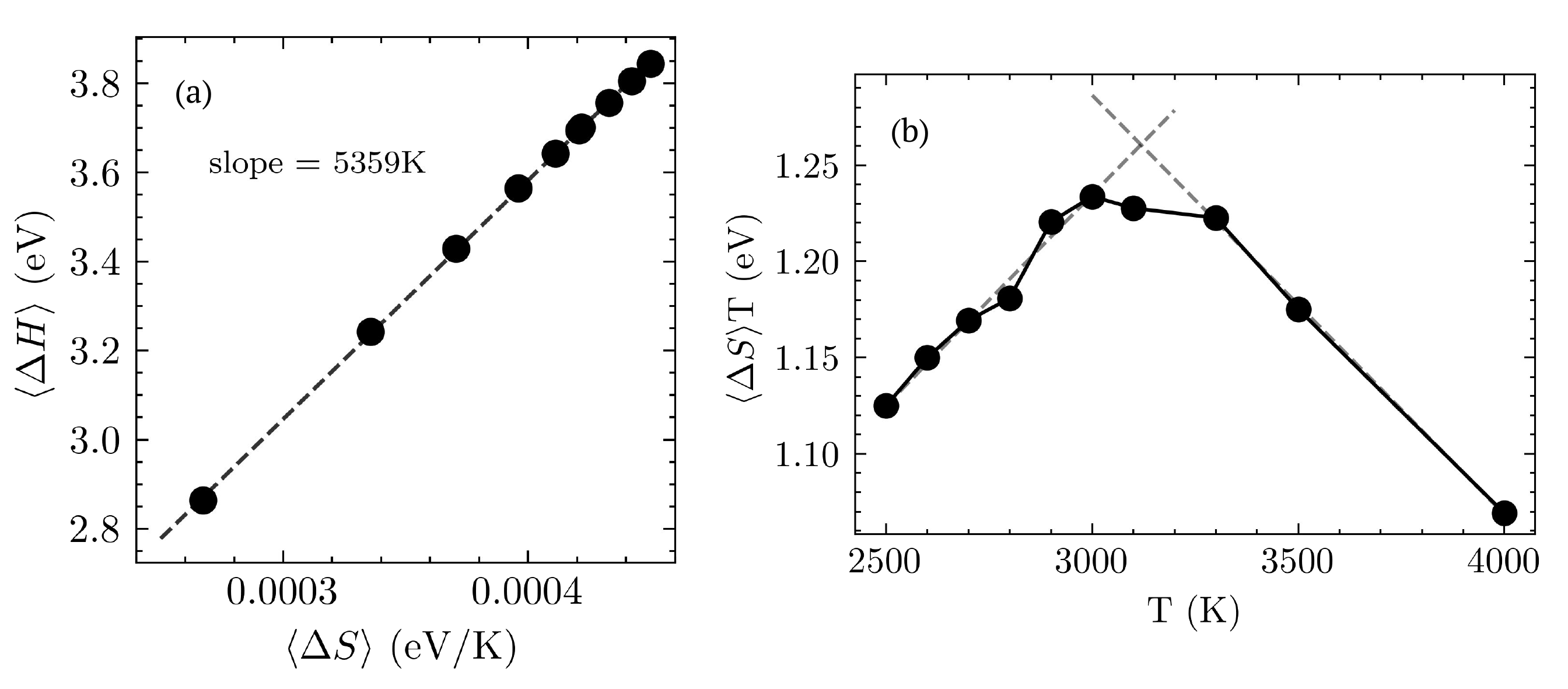}
	\caption{(a) A linear relationship between average activation enthalpy and activation entropy of microscopic dynamics. The slope of 5359 K is an onset temperature where
		diffusion of Si atoms in silica becomes independent of the local atomic structure, i.e., the upper bound temperature of the PEL-influenced region (see main text). (b) A break of slope at 3100 K can be more clearly observed in the entropic contribution of activation energy corresponding to the FTS.}
	\label{fig:delta_S}
\end{figure*}

As the other factor in the elementary rearrangement probability (Equation \ref{eq:eq2}), the activation entropy is approximately linear to the average activation enthalpy, as shown in Fig. \ref{fig:delta_S}a. This relationship given by $\Delta H=T_\mathrm{onset} \Delta S+\Delta H_0$ is consistent with the enthalpy-entropy compensation observed in many systems.\cite{almondActivationEntropyTransport1987} The slope $T_\mathrm{onset}\sim$ 5360 K indicates an onset temperature where diffusion of Si atoms in silica is independent of the local atomic structure, separating the activated dynamics regime for supercooled liquids and the free diffusion regime of high-T liquids. It is consistent with the temperature where inherent enthalpy $e_\mathrm{IS}$ starts to show the typical decreasing trend with temperature during melt-quenching simulations of BKS silica.\cite{saika-voivodFragiletostrongTransitionPolyamorphism2001,sastrySignaturesDistinctDynamical1998} When the entropic contribution to the activation energy is plotted versus temperature, as shown in Figure \ref{fig:delta_S}b, a break of slope is again evident at $\sim$3100 K, separating the fragile and strong regions. It is worth mentioning that, calculating the bulk diffusivity by extrapolating the softness distribution in Fig. \ref{fig:distributions}b to the experimental glass transition temperature (detailed in SI Sec. S4) results in a fairly good agreement with the experimental data, as shown in the inset of Fig \ref{fig:D}.\cite{brebecDiffusionSiliciumDans1980a}

\section{Discussion}\label{sec12}

To summarize, we demonstrate that the FTS in silica liquid can be explained by changes in the microscopic dynamics. Specifically, the energy barriers for local rearrangements in silica liquid show a bimodal distribution representing two distinct microscopic channels with very different barrier heights. The microscopic dynamics channel associated with very small energy barriers ($\Delta H\sim0$ eV), or ``barrierless", closes rapidly as the temperature decreases, while the activation energetics of the channel with larger barriers ($\Delta H\sim4$ eV) has a much weaker temperature dependence. This consequently leads to a crossover in the overall activation energetics at around 3100 K, which manifests itself in the bulk diffusive dynamics as the FTS. Interestingly, water, which is known to have a FTS crossover, is also speculated to have similar bimodality of local structures.\cite{shiOriginEmergentFragiletostrong2018}

\begin{figure*}
	\centering
	\includegraphics[width=1.0\linewidth]{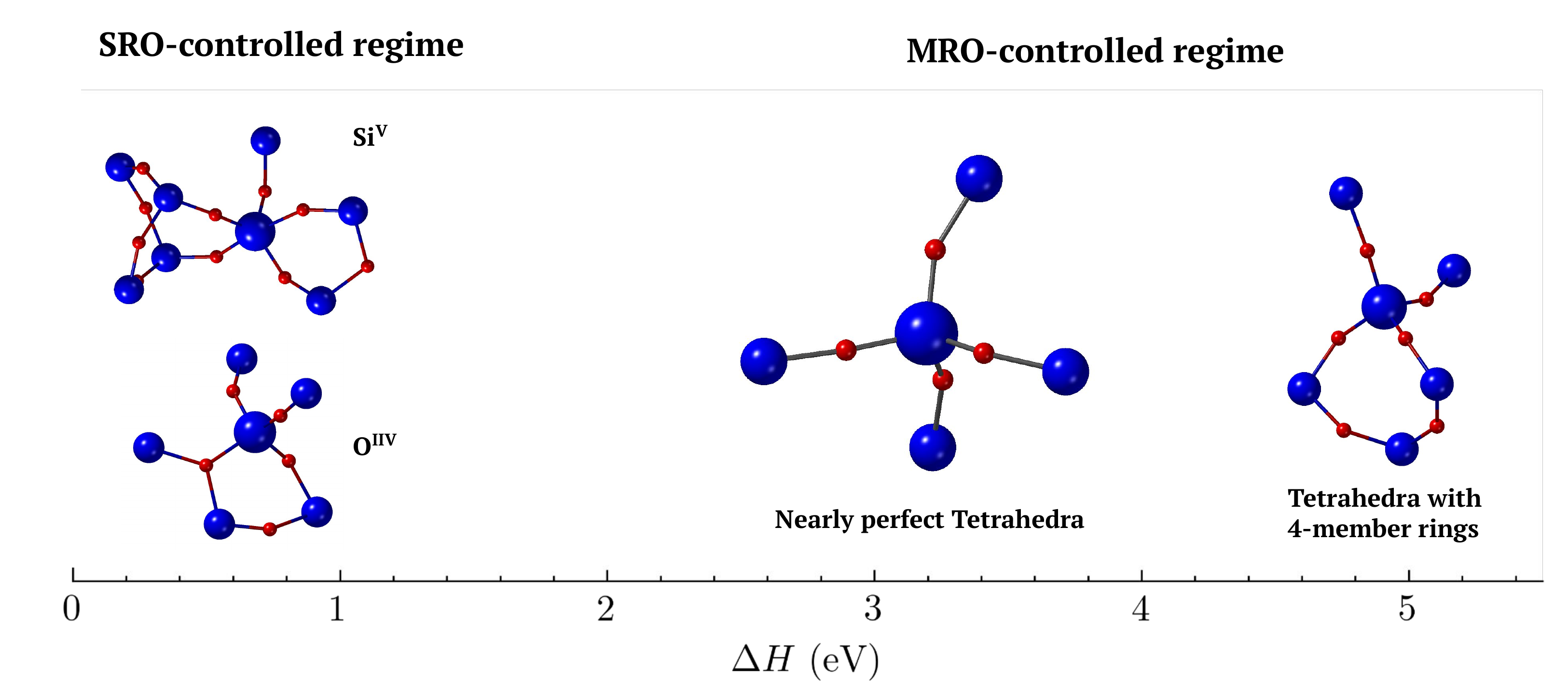}
	\caption{Example local structures of Si atoms with different activation enthalpies or energy barriers of microscopic dynamics. The small energy barriers ($<$1 eV) are associated with defects in the short-range order (SRO), including coordination defects and 3-member rings. The large energy barriers having converged SRO are controlled by the medium-range order (MRO). The most common structures associated with $\sim$2-4.5 eV barriers are near-ideal SiO\textsubscript{4} and SiSi\textsubscript{4} tetrahedra. A less common structure associated with large barriers ($>$4.5 eV) involving distorted tetrahedra with 4-member rings is also shown. }
	\label{fig:examples}
\end{figure*}

By using physically meaningful inputs for machine learning, we can directly investigate the structural features correlating with the two microscopic dynamics channels (see SI Sec. S5 for details). The smaller energy barriers have strong correlation with defects in the short-range order (SRO), including coordination defects, 3-member rings, and highly distorted SiO$_4$ tetrahedra. Typical structures of these short-range defects are shown in Fig. \ref{fig:examples}. These structures have strong tendency to rearrange in a short time. The larger energy barriers, on the other hand, are mostly associated with structures with near-ideal SRO, i.e., SiO$_4$ tetrahedra with small distortions. The medium-range order (MRO) for these structures is usually well established as well, in the form of SiSi$_4$ tetrahedron. Small deviations in the SRO and MRO from the ideal tetrahedra do lead to some variations in the softness. It is interesting to note that we observe the barriers above 4.5 eV are associated with a specific MRO that involves distorted SiSi$_4$ tetrahedra involving 4-member rings, suggesting this MRO structure has strong kinetic stability. Although this MRO structure only occurs with a small fraction of the Si atoms, how it is related to silica stability may warrant more investigations. 

The link between fragile behavior and SRO defects has been noted in the previous study by Saksaengwijit et al..\cite{saksaengwijitOriginFragiletoStrongCrossover2004} Here, we propose that the fragile behavior above FTS can be understood by considering the formation of SRO defects as thermally activated like the formation of intrinsic defects in crystals. Previous studies have already shown the concentration of certain SRO defects in silica has strong dependence on temperature (or inherent enthalpy).\cite{vollmayrCoolingrateEffectsAmorphous1996,yuStructuralSignaturesThermodynamic2021} Some of these defects, such as under or over coordinated Si, are structurally similar to vacancies and interstitials in quartz.\cite{wangNatureRadiationinducedDefects2015} However, unlike in crystals where point defects and their formation energy are usually well defined, the SRO defects in glass structures are complex and may involve different levels and types of distortions without obvious coordination defects. This is evident in Fig. \ref{fig:distributions}a where a broad range of large softnesses are observed for atoms in liquid silica. Nonetheless, all these SRO defects can be associated with small microscopic activation energies. As their concentrations quadratically increase with temperatures, the average activation energy decreases, leading to super-Arrhenius behavior in dynamics. Based on this explanation, FTS crossover should be a universal feature of strong glass-forming liquids. Strong behavior is an indication that there exist one or more diffusion channels in the liquid, of which the activation energetics shows weak temperature dependence. In these liquids, the formation of SRO defects will open additional diffusion channels associated with low activation barriers. The crossover occurs when the  defect concentration begins to strongly affect the average activation energetics. This is echoed by the recent study on various doped silica melts.\cite{mauroViscositySilicaDoped2022} For fragile liquids, however, the universality of FTS crossover upon cooling is unclear because there is no guarantee that all liquids have diffusion channels with weakly temperature-dependent energetics. 

The insights into the FTS from the microscopic dynamics perspective also suggests that the fragile behavior does not require the presence of cooperative motion. In this study, the fragile behavior in the silica bulk diffusivity is derived from microscopic dynamics under the Stokes-Einstein relationship. This lends support to the important role of microscopic dynamics in glass dynamics in the recent debate.\cite{wyartDoesGrowingStatic2017,berthierCanGlassTransition2019} It would be interesting to investigate common fragile liquids to see if there also exist multiple microscopic dynamics channels that show different temperature dependencies. These could be due to SRO defects but can also involve different molecular motions for organic glasses. For studying defects specifically, the softness-based ML approach addresses a major challenge in defining SRO defects in complex amorphous structures.

Finally, question remains on the connection between the origins of the FTS in microscopic dynamics and in configurational entropy. Like in the previous study, our simulations also show FTS is associated with a inflection point in the $S_{c}$ vs. T (see SI), suggesting there may be fundamental correlations between $S_{c}$ and microscopic dynamics. This connection was also suggested recently by Berthier et al. to explain the efficiency of the SWAP algorithm within the RFOT framework.\cite{berthierCanGlassTransition2019} The SWAP algorithm, which enhances microscopic dynamics in Monte Carlo simulations by introducing an additional degree of freedom associated with particle sizes,\cite{ninarelloModelsAlgorithmsNext2017} is able to drastically accelerate glass relaxation, a process that has been explained solely based on thermodynamics in theories like RFOT.\cite{lubchenkoTheoryStructuralGlasses2007,berthierTheoreticalPerspectiveGlass2011} This was explained by that the effective energy landscape (i.e., the energy landscape seen by the system) is altered by artificial enhancements in the microscopic dynamics, circumventing the  metastability that would prevent the system from efficient relaxation in the absence of SWAP. This concept is similar to the concept of ergodicity.\cite{mauroStatisticalMechanicsGlass2014} Here, by demonstrating the microscopic dynamics origin of the FTS in silica, we provide yet another evidence that quantitatively incorporating the effect of microscopic dynamics on broken ergodicity or effective energy landscape is a critical step towards a full thermodynamic description of supercooled liquids.

\section{Methods}\label{sec:M}

\subsection{Machine learning methods}\label{sec:methods_ML}
In this study, the ML model is trained to predict rearrangements of individual atoms in silica liquids. The input features are local structural features of atoms. Previous studies and our test studies show that short-range features within 4 {\AA} in silica are the most predictive for rearrangement. Structures within this range involve SiO\textsubscript{4} and SiSi\textsubscript{4} tetrahedra. Therefore, we use structural features based on these two types of tetrahedra instead of a large number of symmetry functions as in previous studies. For Si atoms specifically, the 10 input features are numbers of O around Si in the first neighboring shell, means and variances of the Si-O bond lengths, means and variances of the O-Si-O bond angles, numbers of Si around center Si in the first neighboring shell, means and variances of the Si-Si distances, means and variances of the Si-Si(center)-Si angles. The 10 features quantitatively describe how the two types of tetrahedra are distorted from the ideal geometry, and atoms in more distorted tetrahedra are expected to rearrange more easily. 

The output of the model is whether an atom will rearrange in the next 1 ps (4 ps was used in the previous study for silica \cite{cubukUnifyingFrameworkStrong2020}). We find that the accuracy of the ML model improves as the time window becomes shorter, although the computational cost for preparing and analyzing data also increases significantly. Besides the previously used hopping probability $p_\mathrm{hop}$ (an indicator of atomic displacements within a given time window),\cite{candelierSpatiotemporalHierarchyRelaxation2010,smessaertDistributionLocalRelaxation2013} we also employ local connectivity changes to identify rearrangements. Connectivity is an important feature of glass formers with a rigid network like SiO\textsubscript{2}. We find that, although the accuracy at the training temperature is not strongly affected, using connectivity helps the ML model make more accurate predictions at higher temperatures. This is because connectivity can effectively distinguish hops from vibrational motions. The cutoff to calculate Si-O connectivity is set to 2 {\AA} based on the range of the first peak in the pair distribution function.

The dataset used for training and testing (25\%) are 13,000 inherent structures from MD trajectories at 2600 K, the lowest temperature investigated in this study. Training the ML model at a low temperature allows the model to better capture the effect of structure on the dynamics. However, rearranging atoms are rare comparing to non-rearranging atoms at this temperature. To avoid naive solutions due to class imbalance, the dataset contains the same number of rearranging and non-rearranging events. This is achieved by randomly select a subset of non-rearranging atoms to match the number of rearranging atoms captured from the simulation. 

For ML, we use logistic regression with $l_2$ regularization in this study.\cite{lecessieRidgeEstimatorsLogistic1992} In our tests, its performance is as good as non-linear classification methods including neural network, random forest, and previously used support vector machine. Similar to previous studies, softness herein is defined to be proportional to the distance to the hyperplane in the feature space.\cite{schoenholzStructuralApproachRelaxation2016} The hyperplane is an $n-1$ dimensional subspace in an $n$ dimensional feature space that best separates instances into two classes. Therefore, softness quantifies the probability of one instance to be classified into one class, that is, rearrangement probability (during 1 ps in our case). Note that the quantity of "softness" would have different mathematical definitions when different ML algorithms are employed. In logistic regression based classification, softness $s$ has a simple form $s=w^\top x+b$, which is a linear combination of input features with a bias.

\subsection{Molecular dynamics simulations}

The systems contain 1512 Si atoms and 3024 O atoms in a cubic simulation box with periodic conditions applied in all three directions. The box size is fixed in the canonical (NVT) ensemble to maintain a density of 2.28 g/cm\textsuperscript{3}. The time step is 1 fs and the BKS potential is employed in all the simulations. The cutoff of the potential is chosen as 6.0 {\AA}, which can reproduce the experimental density of 2.2 g/cm\textsuperscript{3} if the system is melt-quenched under isothermal-isobaric (NPT) ensemble. The temperature range investigated covers 2600 to 4000 K. In simulations at a specific temperature, the initial structure is taken at the same temperature from a melt-quenching simulation trajectory with a cooling rate of 0.01 K/ps. To make sure equilibrium is reached, the system is annealed at each temperature for a sufficiently long time with respect to the relaxation time calculated the from intermediate scattering function (e.g., the system was annealed for $>$200 ns at 2600 K, comparing with the relaxation time of $\sim$50 ns, as shown in Fig. S5). To obtain inherent structures, energy minimization is performed using an MD-based approach that instantaneously quenches the structure to 0 K under 1 bar pressure. The inherent structure is then obtained after continuing the simulation at 0 K for 10 ps. We find this is sufficient to converge the change in the system energy to below 10$^{-4}$ meV per atom in all cases. 

% \backmatter

% \bmhead{Supplementary information}

% Supplementary information of this article is provided as a separate document. 

\begin{acknowledgments}
This research was primarily supported by NSF through the University of Wisconsin Materials Research Science and Engineering Center (DMR-1720415).This work used the Extreme Science and Engineering Discovery Environment (XSEDE), which is supported by National Science Foundation grant number ACI-1548562.
\end{acknowledgments}

\bibliography{FTS}

\end{document}

\title[]{Supplementary Information for ``Understanding the fragile-to-strong transition in silica from microscopic dynamics''}

\author{Zheng Yu}
\affiliation{Department of Materials Science and Engineering, University of Wisconsin-Madison, Madison, 53706, USA}
%  \altaffiliation[Also at ]{Physics Department, XYZ University.}%Lines break automatically or can be forced with \\
\author{Dane Morgan}
\affiliation{Department of Materials Science and Engineering, University of Wisconsin-Madison, Madison, 53706, USA}

\author{M. D. Ediger}
\affiliation{Department of Chemistry, University of Wisconsin-Madison, Madison, 53706, USA}

\author{Bu Wang}%
 \email{bu.wang@wisc.edu}
\affiliation{Department of Materials Science and Engineering, University of Wisconsin-Madison, Madison, 53706, USA}
\affiliation{%
 Department of Civil and Environmental Engineering, University of Wisconsin-Madison, Madison, 53706, USA
}

% % % \equalcont{These authors contributed equally to this work.}

\maketitle

\tableofcontents

\section{Configurational entropy}

As introduced in the main text, the fragile-to-strong transition (FTS) in silica has been so far explained based on configurational entropy. In this work, we recalculate the configurational entropy of silica based on our replica exchange molecular dynamics (REMD) simulations to confirm this explanation. We apply the method previously used for calculating configurational entropy calculations,\cite{sciortinoInherentStructureEntropy1999} with the formalism proposed by Stillinger and Weber.\cite{stillingerPackingStructuresTransitions1984} The basic idea is to count the number of inherent structures apply the following equation
\begin{equation}
    S_\mathrm{conf}(e_\mathrm{IS})/k_B = \log[P(e_\mathrm{IS},T)] + \beta e_\mathrm{IS} + \beta f(\beta,e_\mathrm{IS})
    \label{eq:eq1}
\end{equation}
where $e_\mathrm{IS}$ is the potential energy or enthalpy of the inherent structure, $P(e_\mathrm{IS},T)$ is the probability of sampling $e_\mathrm{IS}$ at $T$, $f(\beta,e_\mathrm{IS})$ is the free energy of the system confined in one energy well with $e_\mathrm{IS}$, and $\beta=1/k_BT$. The method assumes that the energy wells on the potential energy landscape (PEL) have similar shapes independent of $e_\mathrm{IS}$, therefore Equation \ref{eq:eq1} can be superimposed at different temperatures by shifting $\beta f(\beta)$. 

\begin{figure}[H]
  \centering
  \includegraphics[width=0.95\linewidth]{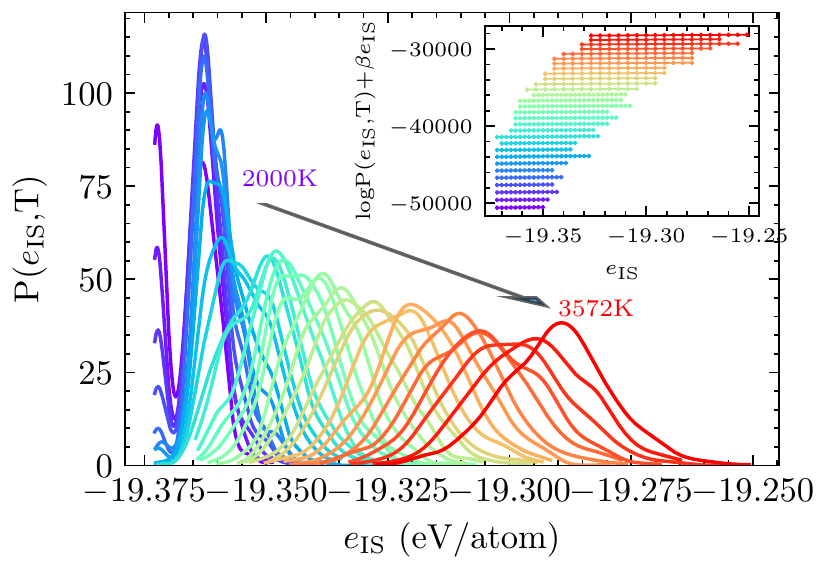}
  \caption{Probability distributions of energies of inherent structures $e_\mathrm{IS}$ in REMD simulations at 30 temperatures from 2000 to 3572 K. The distributions at the lowest 6 temperatures below 2300 K show narrower main peaks and secondary peaks due to not completely reaching equilibrium. The inset plots the intermediate product for calculating configurational entropy as in Equation \ref{eq:eq1}.}
  \label{fig:Sc_calc}
\end{figure}

In our REMD simulations, we run replicas in the isothermal-isobaric (NPT) ensemble at 30 temperatures between 2000 K and 3572 K. The temperature range is selected to cover the fictive temperature of silica identified in the melt-quenching simulations and the melting point. For simulation efficiency, the 30 temperatures are distributed to yield similar exchange acceptance rates between all the neighboring temperatures (with a gradually increasing temperature interval). The inherent structures are obtained from snapshots of trajectories by energy minimization. The simulation system contains 150 Si atoms and 300 O atoms in cubic simulation boxes. The BKS potential is employed for all the simulations with a cutoff of 8.5 \AA. More details of the simulations can be found in reference.\cite{yuStructuralSignaturesThermodynamic2021}  

\begin{figure}[]
  \centering
  \includegraphics[width=0.95\linewidth]{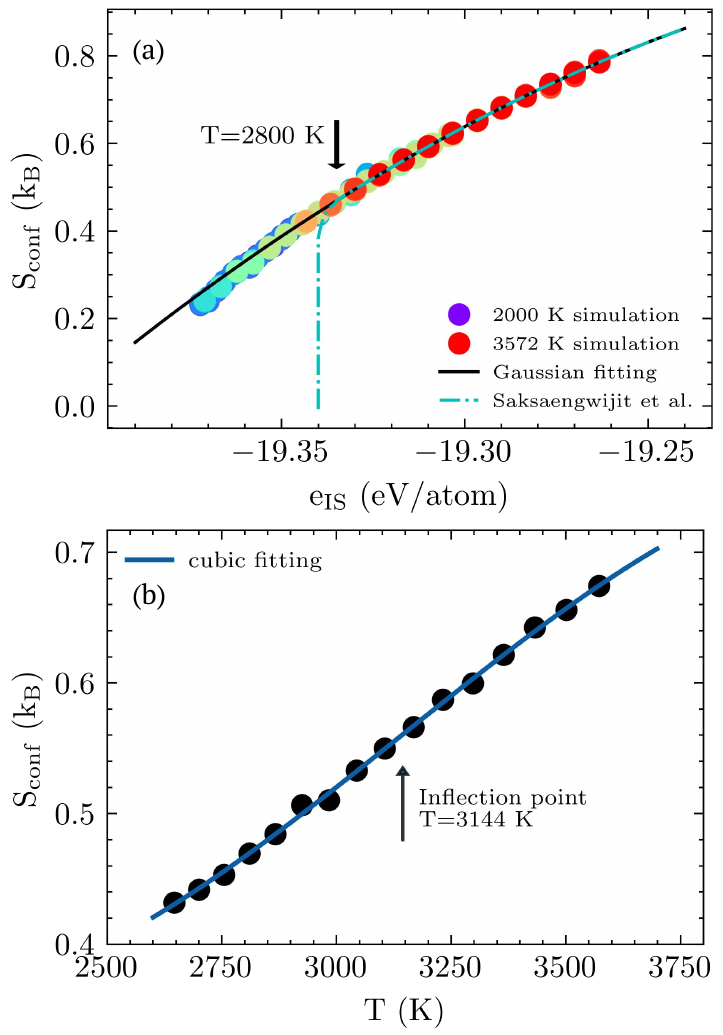}
  \caption{Configurational entropy of silica as a function of (a) energy of inherent structure $e_\mathrm{IS}$ and (b) temperature, calculated based on REMD simulation results at temperatures from 2000 K to 3572 K. The solid line in (a) is a quadratic fitting using data from only high temperatures, and the line in (b) is a cubic fitting using data from all the temperatures. The inflection point in (b) explains the FTS transition based on the Adam-Gibbs relation, confirming the argument from Saika-Voivod et al..\cite{saika-voivodFreeEnergyConfigurational2004}  However, the depletion of states in silica PEL proposed by Saksaengwijit et al. as indicated by the dashed point line in (a), is not supported by our data.\cite{saksaengwijitOriginFragiletoStrongCrossover2004}.
}
  \label{fig:Sc}
\end{figure}

The probability distributions of energies of inherent structures $e_\mathrm{IS}$ at the simulated temperatures are shown in Fig. \ref{fig:Sc_calc} and the results before shifting $\beta f(\beta)$ are shown in the inset. Note that the distributions at the lowest 6 temperatures below 2300 K show narrower main peaks and secondary peaks because equilibrium is not reached. From these distributions, we can calculate the relative change of configurational entropy with temperature based on Equation \ref{eq:eq1}. The absolute reference used here is the same as in the previous work.\cite{saika-voivodFragiletostrongTransitionPolyamorphism2001}. Figure \ref{fig:Sc} shows the calculated configurational entropy of silica as a function of $e_\mathrm{IS}$ and temperature. The depletion of states in silica PEL proposed by Saksaengwijit et al., plotted as the dashed point line in (a), is not supported by our data.\cite{saksaengwijitOriginFragiletoStrongCrossover2004}  However, the $S_\mathrm{conf}$ inflection point in (b) does explain the FTS transition based on the Adam-Gibbs relation, confirming the explanation by Saika-Voivod \emph{et al.}.\cite{saika-voivodFreeEnergyConfigurational2004} Note that the calculated $S_\mathrm{conf}$ in (a) at the lowest temperatures are slightly lower than the fitting curve based on high temperature data, which again is a result of the insufficient sampling not reaching equilibrium.

\section{Machine learning results}

\begin{figure}[]
  \centering
  \includegraphics[width=1.0\linewidth]{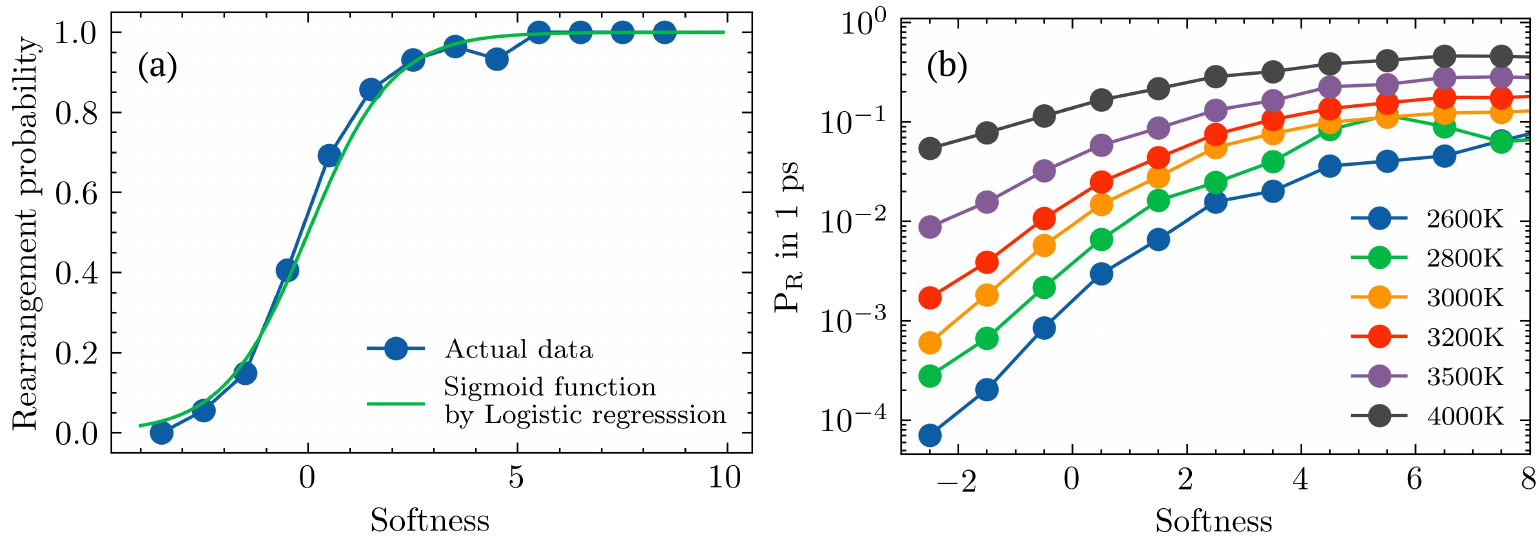}
  \caption{(a) Rearrangement probability as a function of softness in the learning dataset (ratio=50/50). The actual data points are in good agreements with the sigmoid function from the logistic regression model, suggesting a good performance of the trained model.  (b) Rearrangement probability of individual atoms in 1 ps as a function of softness in the raw dataset (without manipulating the ratio between the numbers of rearranging vs. non-rearranging atoms).}
  \label{fig:ML_results}
\end{figure}

In Fig. \ref{fig:ML_results}a, we show the actual rearrangement probability of individual Si atoms within 1 ps as a function of softness in the machine learning dataset. The data points are in good agreements with the sigmoid function $P_R=1/(1+\exp(-s))$ from the logistic regression model, confirming the training has achieved a good performance. The accuracy of the ML model on the test data is 86\% (precision: 93\%, recall: 79\%, F1:85\%). Figure \ref{fig:ML_results}b plots the rearrangement probability as a function of softness at various temperatures. Note that these are different from Fig. \ref{fig:ML_results}a because the fractions of two classes (rearranging vs. non-rearranging) are no longer manipulated to be 50/50. 

\section{Connecting short-term rearrangements to long-term dynamics}

\begin{figure*}[]
  \centering
  \includegraphics[width=1.0\linewidth]{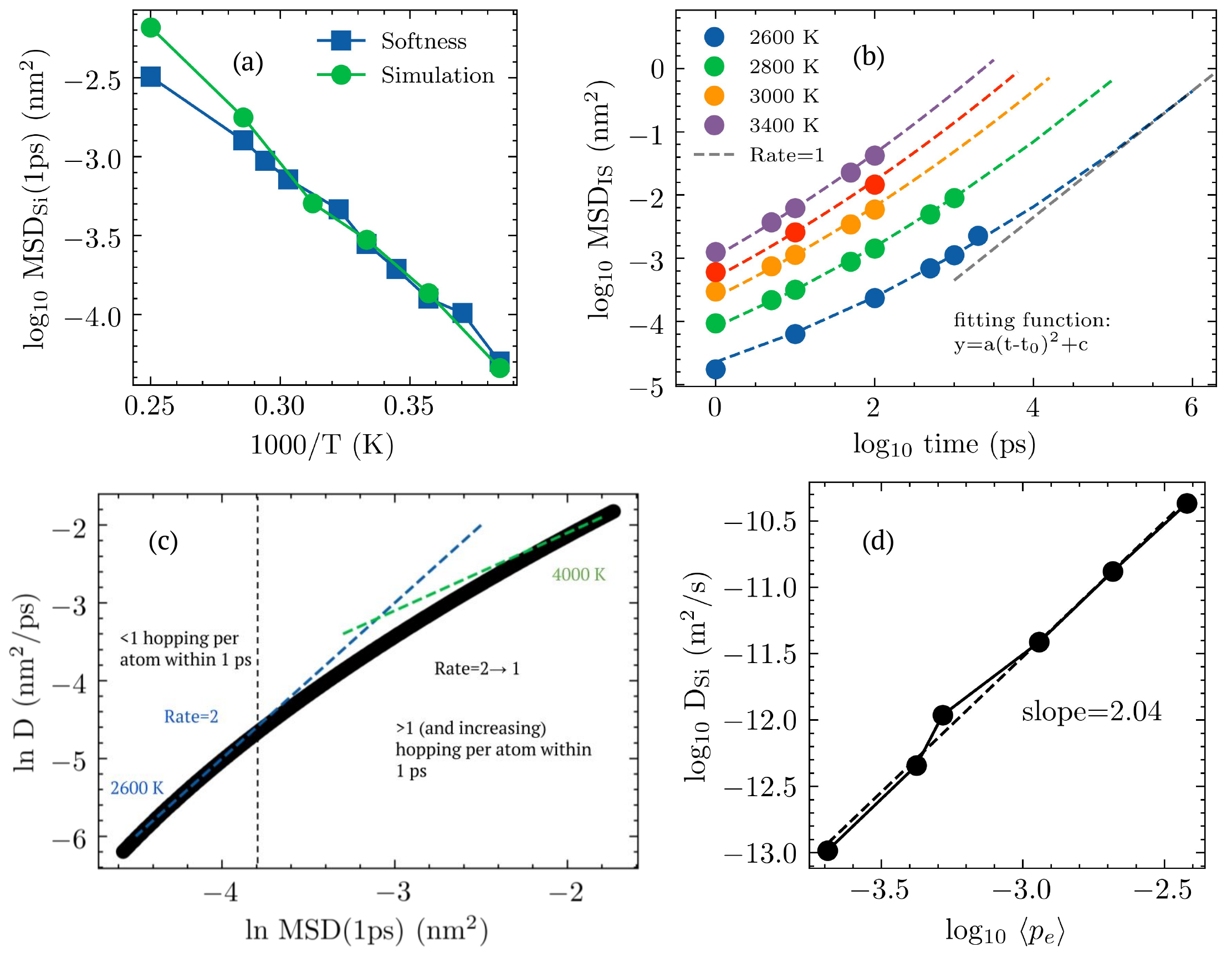}
  \caption{(a) Mean squared displacements based on inherent structures for Si atoms in silica predicted by softness compared to those directly calculated
 from MD simulations. (b) The same quantity as in (a) as a function of time at various temperatures. The fitting parameter $t_0$ is an approximate linear function of the inverse of temperature. (c) and (d), diffusion coefficients of Si atoms in silica versus the corresponding inherent structure MSD within 1 ps, and the elementary hopping probability (both in log-log scale), respectively. The slope of the curves in (c) at low temperatures and in (d) is around 2, suggesting that long-term and short-term dynamics are approximately connected by a quadratic function.}
  \label{fig:short2long}
\end{figure*}

As stated in the main text, if assuming a constant hopping distance $R$, the short-term mean square displacement (MSD) based on inherent structures is proportional to the average hopping probability as  
\begin{equation}
    \mathrm{MSD}_{\mathrm{IS}}(t<<\tau_{\alpha}) \propto \langle R^2 \exp(-\frac{\Delta G}{k_B T}) \rangle_{eq} \propto R^2 \langle \exp (-\frac{\Delta G}{k_B T}) \rangle_{eq} 
\end{equation}
As shown in Fig. \ref{fig:short2long}a, MSD$_\mathrm{IS}$(1ps) predicted by softness agree well with those calculated directly from simulations at lower temperatures where hopping rarely occurs within 1 ps. The softness predicted MSD becomes become increasingly underestimated as the temperature increases above 3500 K, because at high temperatures hoppings are more frequent and diffusive dynamics could occur within 1 ps.

\begin{figure}[]
  \centering
  \includegraphics[width=0.95\linewidth]{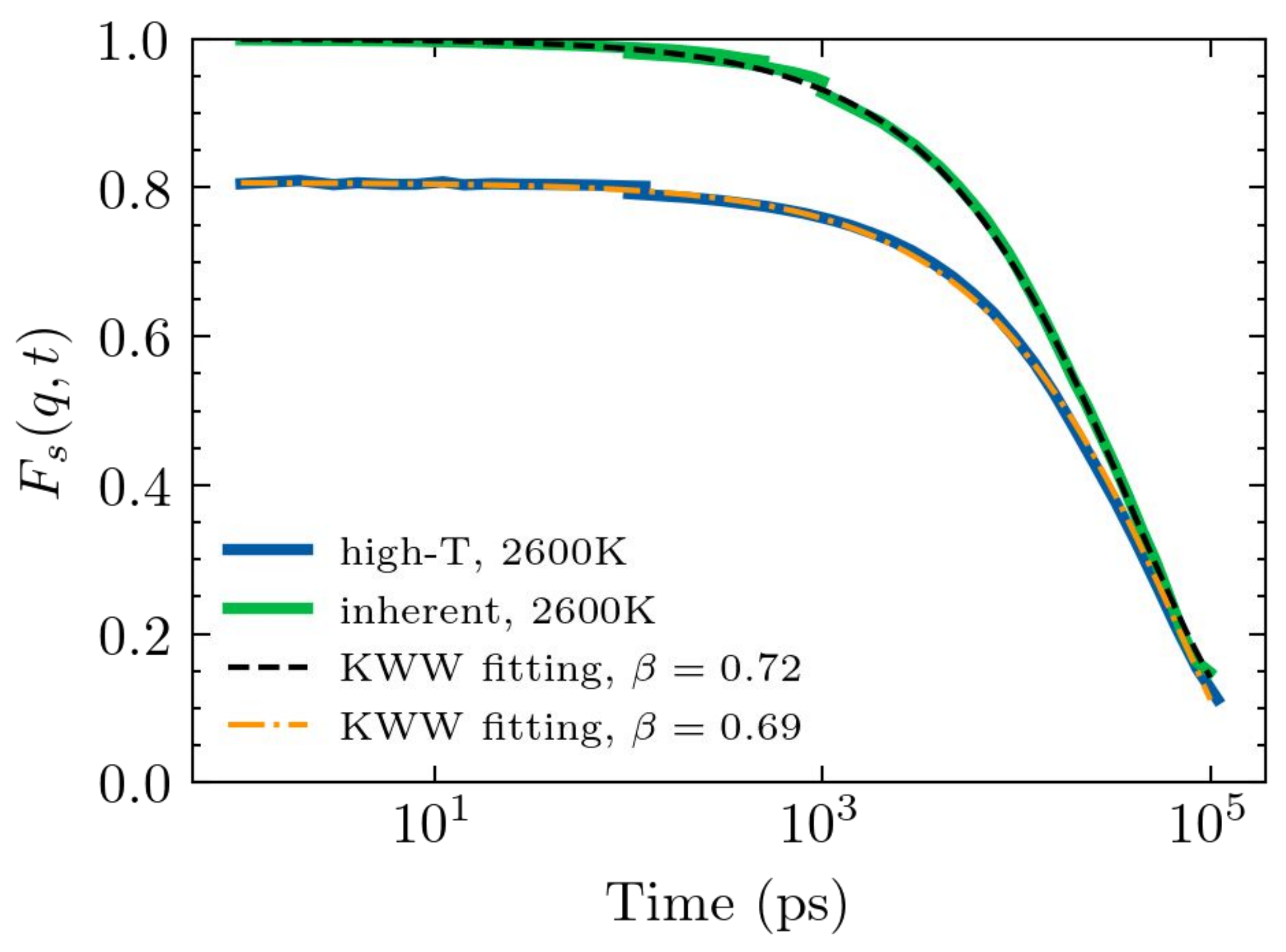}
  \caption{Intermediate scattering functions $F_s$ in silica at 2600 K. Comparing with $F_s$ calculated using snapshots of high-temperature MD trajectory (labeled as high-T), $F_s$ calculated based on inherent structures (labeled as inherent) does not show any plateau or two-step relaxation, suggesting vibrations are excluded and only one relaxation mode remains in silica.}
  \label{fig:scattering}
\end{figure}

To investigate the connection between microscopic dynamics and long-term diffusion, it is necessary to exclude thermal vibrations from dynamics calculations. Therefore, we use MSD based on inherent structures (MSD$_\mathrm{IS}$) to quantify dynamics. Previously, the Monte Carlo simulation was used to exclude part of atomic vibrations, but it does not exclude cluster vibration, which shows up as the plateau corresponding to the two-step relaxation in the self-intermediate scattering function.\cite{berthierRevisitingSlowDynamics2007} Here in silica, the plateau in the self-intermediate scattering function disappears when inherent structures are used, as shown in Fig. \ref{fig:scattering}. This also suggests there is no ``$\beta$" relaxation modes in silica, so all the local rearrangements are in some way connected to the $\alpha$ relaxation only.

Figure \ref{fig:short2long}b shows MSD$_\mathrm{IS}$ of Si atoms in silica as a function of time at various temperatures. The slope of the curves gradually increase from 1 ps ($<$1) up to the Fickian region (=1), which can be most clearly seen at lower temperatures. We find that these curves in silica can be empirically described by quadratic function $y=a(t-t_0)^2+c$, where $a$ and $c$ are constants and $t_0$ is a fitting parameter depending on $T$. As a result, the curves from different $T$ can be viewed as the same curve with horizontal shifts, in which the slopes are equal at the same MSD$_\mathrm{IS}$. An evidence of this is that MSD required to enter the Fickian region are always similar at different $T$. The parameter $t_0$ is found approximately linear to the inverse of temperature. Based on this, we can empirically estimate MSD$_\mathrm{IS}$ at long duration at a given $T$ from short-term data.  

To view the dynamical evolution more closely, we plot diffusion coefficient $D$ vs. MSD$_\mathrm{IS}$ within 1 ps in a log-log scale in Fig. \ref{fig:short2long}c. The slope of the curve is around 2 at low temperatures, suggesting that $D$ is approximately proportional to the square of MSD$_\mathrm{IS}$ (1ps) as long as the hopping frequency is small. We already know MSD$_\mathrm{IS}$ (1ps) at low temperatures is proportional to $\langle \exp(-\Delta G/kT)\rangle$. Thus, $D$ is approximately proportional to $\langle \exp(-\Delta G/kT)\rangle^2$, as shown in Fig. \ref{fig:short2long}d. Note that this near-quadratic relationship should be expected at all temperatures, although the slope of the curve further increases with increasing temperature as relaxation times become closer to 1 ps. 

\section{Extrapolation to experimental $T_g$}

\begin{figure}[]
  \centering
  \includegraphics[width=1.0\linewidth]{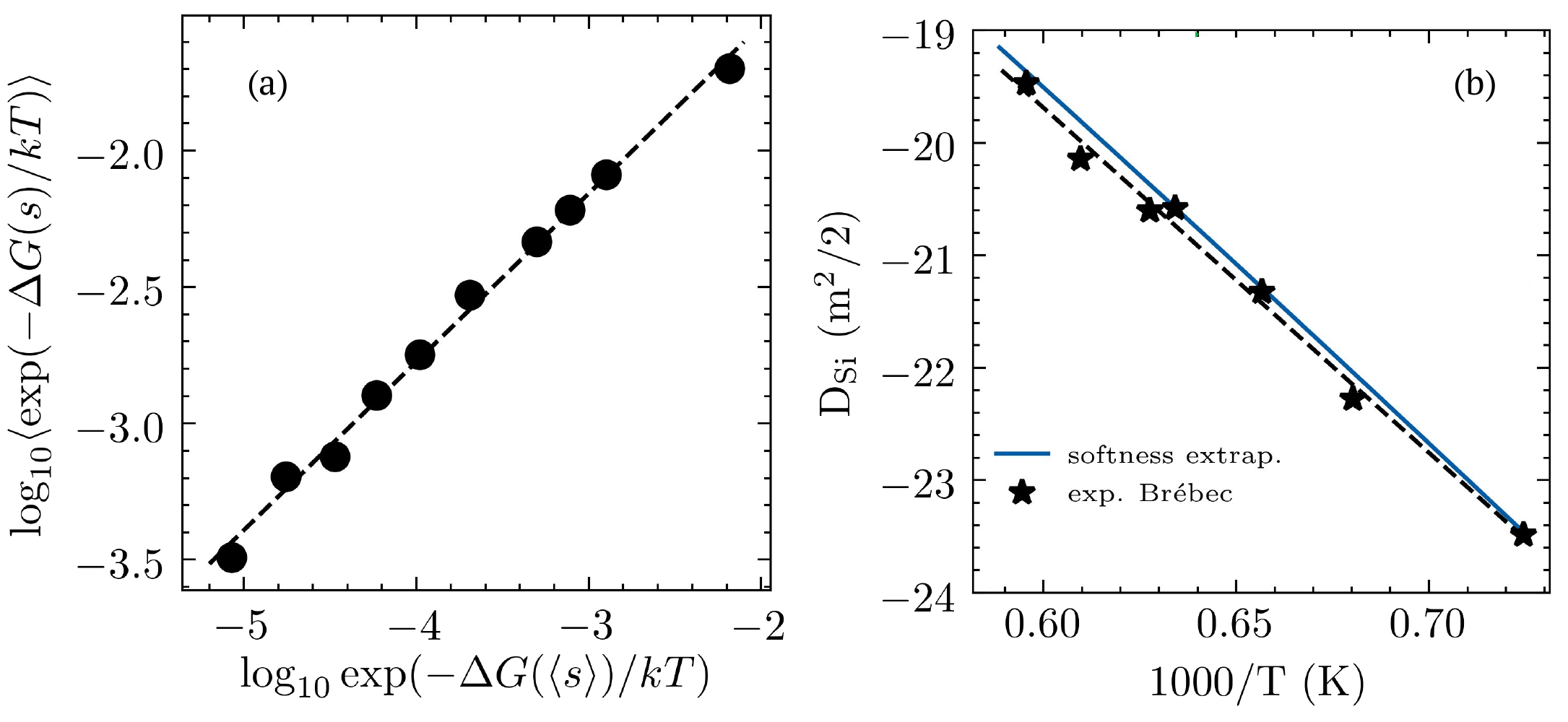}
  \caption{(a) Average elementary hopping probability versus elementary hopping probability calculated using average softness. The linear relationship in the figure suggests $\langle \exp (-G(s)/kT)\rangle \sim  \exp(-0.62*G(\langle s\rangle )/kT)$, resulting from the distribution of elementary energy barriers. (b) Diffusion coefficients of Si atoms in silica estimated by extrapolating softness (also shown in Fig. 3b) to temperatures near the experimental $T_g$ are in good agreement with the experimental values.\cite{brebecDiffusionSiliciumDans1980a}}
  \label{fig:extrapolation}
\end{figure}

As shown in Fig. 3b of the main text, the average softness shows a linear relationship with temperature. We extrapolate this relationship and use the softness to predict diffusion coefficients near the experimental $T_g$. Before the extrapolation, however, we emphasize the difference between $\langle\exp (-G(s)/kT)\rangle$ and $\exp(-G(\langle s\rangle)/kT)$, where $\langle x\rangle$ is thermodynamic average of a system property $x$. As shown in Fig. \ref{fig:extrapolation}a, these two quantities are not equivalent but have a power law correlation (this is due to the specific bimodal distributions of softness in the case of silica). Based on this relation, we can predict diffusion coefficients using the average softness. The total activation energy (the slope of the Arrhenius plot of $D$) is approximately equal to 0.62 times the averaged microscopic activation enthalpy. Figure \ref{fig:extrapolation}b shows the predicted $D$ for Si atoms around the experimental $T_g$. They are in excellent agreement with the experimental data.\cite{brebecDiffusionSiliciumDans1980a}

\section{Local structural features responsible for various energy barriers}

\begin{figure}[]
  \centering
  \includegraphics[width=0.9\linewidth]{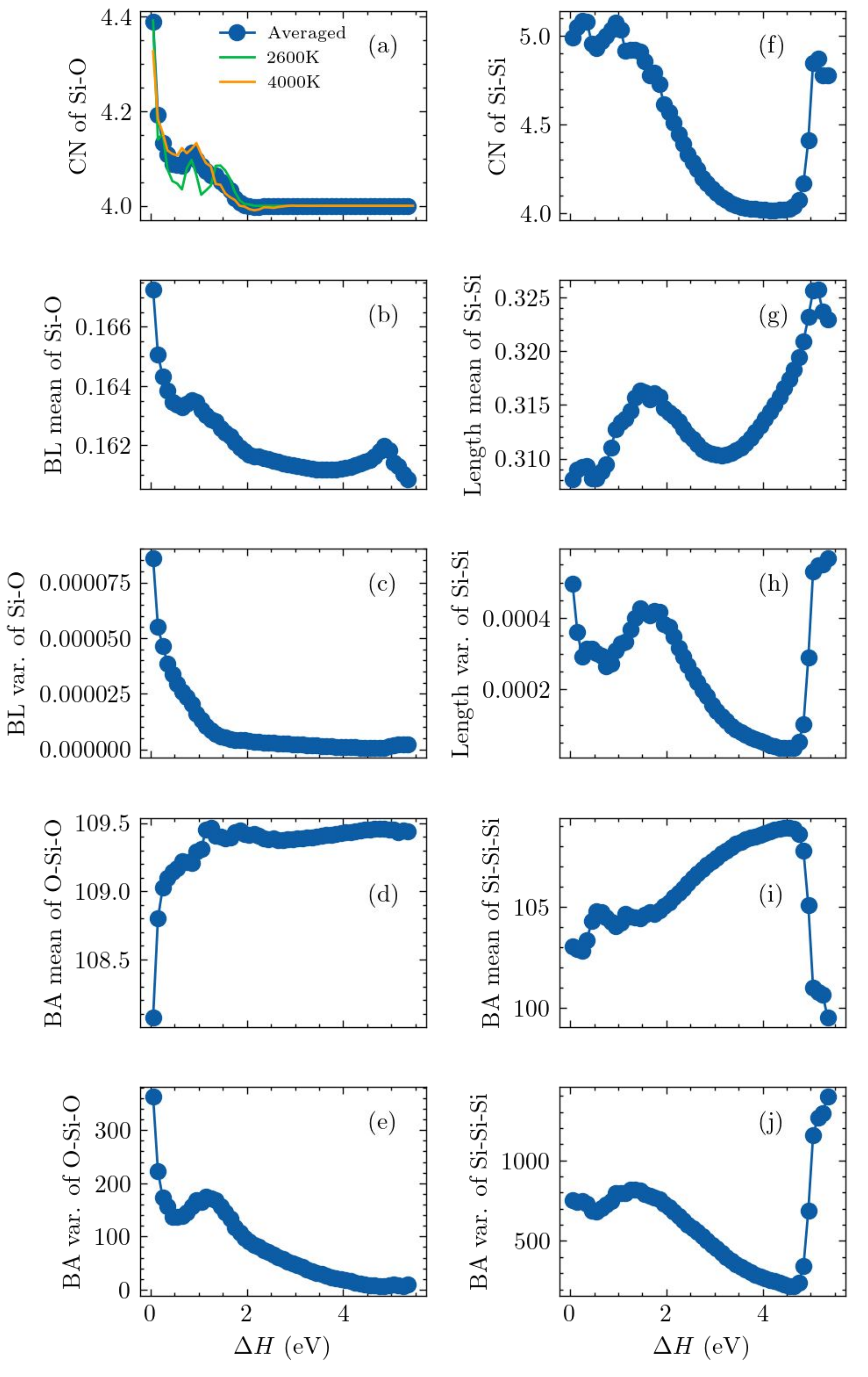}
  \caption{Relationship between the activation enthalpy (or energy barrier) of microscopic dynamics $\Delta H$ and the input structural features: (a-e) coordination number, bond length mean and variance of Si-O, bond angle mean and variance of O-Si-O, based on the SiO\textsubscript{n} polyhedron; (f-j) similar features based on the SiSi\textsubscript{n} polyhedron. The relationships extracted from 4000 K and 2500 K are consistent overall.}
  \label{fig:H_input}
\end{figure}

From the bimodal activation enthalpy distributions shown in Fig. 3c, two types of energy barriers for Si rearrangements are observed in silica. Figure \ref{fig:H_input} shows the relationships between the input local structural features and the activation enthalpy (or energy barrier) of microscopic dynamics $\Delta H$. Note that the relationships extracted from 4000 K and 2500 K are overall consistent, suggesting the activation energetics is successfully attributed to the local structures in the ML model. 

The small energy barriers are related to all Si with coordination defects (number of O around Si $>$4), and also Si in highly distorted tetrahedra. They commonly involve five Si atoms in the first shell of neighboring Si-Si.  The changes of the small energy barriers are more reflected in the SiO\textsubscript{n} polyhedral features (short-range order) than the SiSi\textsubscript{n} ones (medium-range order). Particularly, mean bond length of Si-O and mean bond angle of O-Si-O show the clearest correlation with energy barrier in this region.

Unlike the small energy barriers, the differences between large barriers are more clearly reflected by the SiSi\textsubscript{n} polyhedral features in the medium-range order. As the energy barrier increases up to 4.5 eV, the coordination number of the first Si-Si shell gradually decreases to 4 from 5, the variances of the Si-Si lengths and the Si-Si-Si angles decrease, and the mean of the Si-Si bond lengths increases, suggesting that a Si atom in the center of low-density near-ideal SiO\textsubscript{4} and SiSi\textsubscript{4} tetrahedra has high kinetic stability. More interestingly, higher $\Delta H>$4.5 eV are related to a very different structure. In this new structure, the short-range structure is in ideal tetrahedral order. However, the medium-range structure has five Si atoms in the first Si-Si shell, and the variances of Si-Si lengths and Si-Si-Si angles increase abruptly, to values even larger than those with the small energy barriers. We found this medium range structure is associated with distorted SiSi\textsubscript{4} tetrahedra involving 4-member rings, as shown in Fig. 5. Such a highly kinetically stable structure exists in both low and high temperatures, suggesting it is not a signature of new phases. 

\section{Density of small energy barriers}

\begin{figure}[]
  \centering
  \includegraphics[width=1\linewidth]{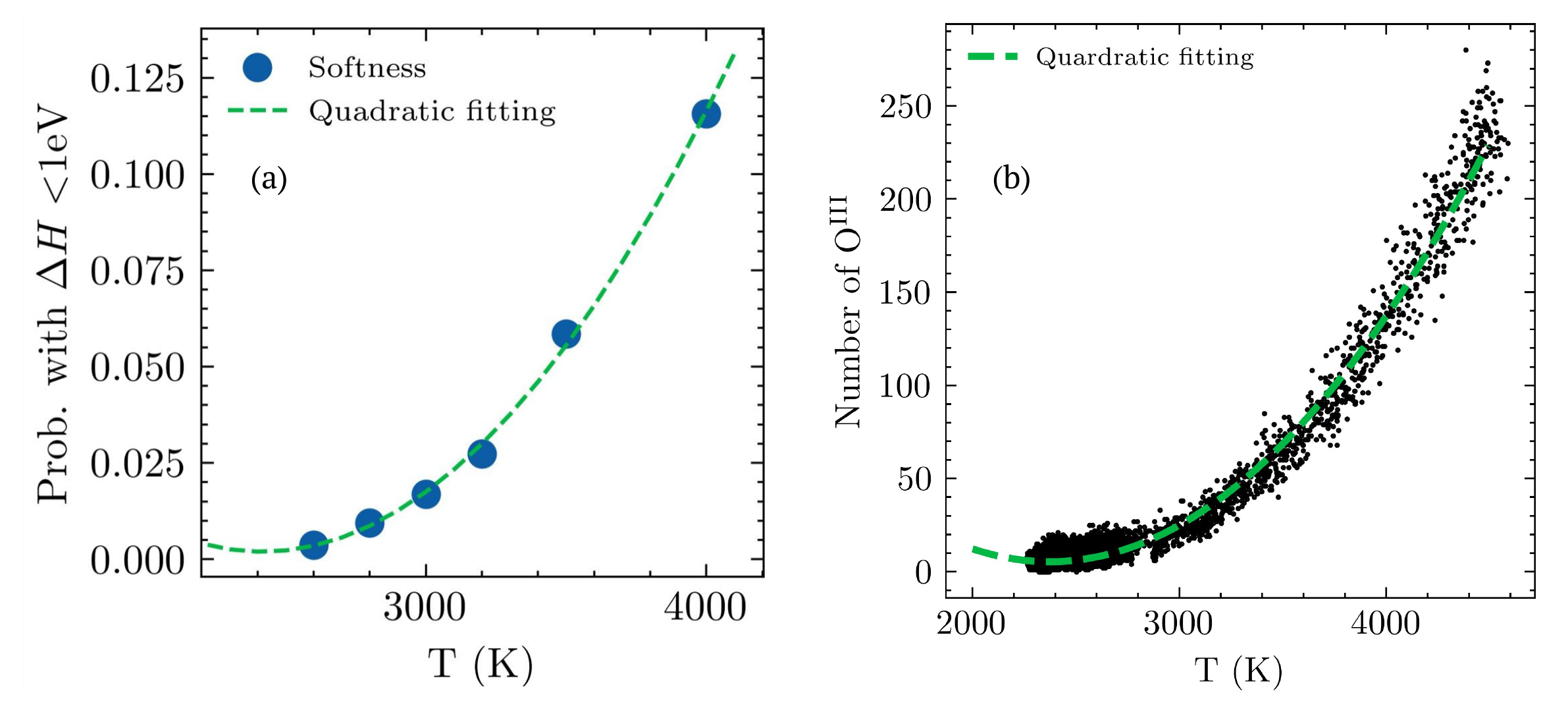}
  \caption{(a) Probability of Si atoms in BKS silica with small $\Delta H<$ 1 eV, as shown in Fig. 3c, as a function of temperature.(b) Number of oxygen point defects with a coordination number of 3 in silica as a function of temperature.  The relationships can be well described by quadratic functions.}
  \label{fig:population_smallH}
\end{figure}

As discussed in the main text, there are two diffusion channels with different energy barriers in BKS silica. Atoms with small energy barriers  (or microscopic activation enthalpy) $\Delta H<$ 1 eV can be considered as short-range-order (SRO) defects in amorphous materials, playing a significant role on silica's dynamic behaviors. The density of these small energy barriers decreases when lowering temperature, as shown in Fig. \ref{fig:population_smallH}a. The temperature dependence within the investigated range is well described by
\begin{equation}
    N(\Delta H<1 \mathrm{eV}) \propto (T-T')^2
\end{equation}
 where $T'$ is a fitting temperature, indicating the complete depletion of small barriers. Note that this kind of quadratic decrease with temperature was also observed in the numbers of point defects in silica, as shown in Fig. \ref{fig:population_smallH}b.\cite{yuStructuralSignaturesThermodynamic2021} This relation has the same form of what RFOT proposed for the density of two-level systems, $\sim (T-T_K)^2$, where $T_K$ is the Kauzmann temperature,\cite{lubchenkoIntrinsicQuantumExcitations2001,lubchenkoTheoryStructuralGlasses2007} suggesting again that microscopic dynamics might be associated with thermodynamics by a consistent (or similar) lengthscale. However, $T'$ as seen in Fig. \ref{fig:population_smallH} is above 2000 K, much higher than the Kauzmann temperature of silica (expecting a low temperature considering strong dynamics of silica). This disagreement could be due to limited simulation conditions (like the BKS potential, or the small simulation box), or some missing modification in the current theory worth further investigation.

\bibliography{FTS}% common bib file
